\pdfoutput=1
\documentclass[aps,prl,nofootinbib,preprintnumbers,amsmath,amssymb,latexsym,array,enumerate,letter,twocolumn]{revtex4}
\usepackage{mathtools}
\usepackage{epsfig}
\usepackage{graphicx}
\usepackage{color}
\usepackage{amssymb}
\usepackage{enumitem}
\usepackage{amsmath}
\usepackage{hyperref}
\usepackage{breakurl}

\makeatletter

\def\be{\begin{equation}}
\def\ee{\end{equation}}
\newcommand{\bea}{\begin{eqnarray}}
\newcommand{\eea}{\end{eqnarray}}

\begin{document}

%\widetext
%\preprint{DESY-22-XXX}

\title{\LARGE{A Cosmic Higgs Collider}}

\author{Bibhushan Shakya}

\affiliation{ \vskip 2mm William H.\ Miller III Department of Physics \& Astronomy, Johns Hopkins University, 3400 N.\ Charles St., Baltimore, MD 21218, USA}

%\affiliation{Deutsches Elektronen-Synchrotron DESY, Notkestr.\,85, 22607 Hamburg, Germany}

\begin{abstract}
This paper examines frameworks and phenomenology of ultrarelativistic Higgs vacuum bubble collisions in a first-order phase transition associated with the Standard Model Higgs field in the early Universe. Such collisions act as a cosmic scale Higgs collider, providing access to energy scales far beyond any temperature reached in our cosmic history, potentially up to the Planck scale. %Such configurations can be realized in several scenarios: quenched electroweak phase transitions, symmetry-restoring phase transitions of the Higgs from a higher minimum during reheating, or transitions out of a metastable Higgs vacuum at large field values. 
This provides a unique opportunity to probe new physics that couples to the Higgs at very high scales, while also enabling novel applications for various cosmological phenomena, opening tremendous opportunities for particle physics and cosmology. As examples, we demonstrate the viability of nonthermal production of ultra-heavy Higgs portal dark matter up to $10^{16}$ GeV (with observable indirect and direct detection signals up to $m_\text{DM}=\mathcal{O}(10)$ TeV), and leptogenesis from the production of GUT scale right-handed neutrinos.

%Collision of Higgs vacuum bubbles in the early Universe.

\end{abstract}

\maketitle

\section{Motivation}

First-order phase transitions (FOPTs), consisting of the nucleation, expansion, and collision of bubbles of a stable vacuum in a  metastable vacuum background, are well-motivated early Universe phenomena that can be naturally realized in various beyond the Standard Model (BSM) frameworks. FOPTs have become one of the most extensively studied topics in recent years as a promising source of stochastic gravitational waves (GWs) from the early Universe \cite{Grojean:2006bp,Caprini:2015zlo,Caprini:2018mtu,Caprini:2019egz,Athron:2023xlk,Kosowsky:1991ua,Kosowsky:1992rz,Kosowsky:1992vn,Kamionkowski:1993fg,Caprini:2007xq,Huber:2008hg,Jinno:2016vai,Jinno:2017fby,Konstandin:2017sat,Cutting:2018tjt,Cutting:2020nla,Lewicki:2020azd,Hindmarsh:2013xza,Hindmarsh:2015qta,Hindmarsh:2017gnf,Cutting:2019zws,Hindmarsh:2016lnk,Hindmarsh:2019phv,Caprini:2009yp,Brandenburg:2017neh,RoperPol:2019wvy,Dahl:2021wyk,Auclair:2022jod,Jinno:2022fom,Inomata:2024rkt}.

It is now well established that in some FOPT configurations, the latent vacuum energy released during the phase transition can get concentrated into the bubble walls, boosting them to ultrarelativistic (UR) speeds (for the purposes of this paper, we take this to mean a Lorentz boost factor $\gamma\geq 10$). The collisions of such energetic bubbles can act as cosmic scale high energy colliders, capable of producing particles with masses or energies far beyond the temperature of the ambient plasma, up to the Planck scale \cite{Watkins:1991zt,Falkowski:2012fb,Shakya:2023kjf,Mansour:2023fwj,Giudice:2024tcp}.  The dynamics and collisions of such UR bubbles have, for instance, been shown to be viable mechanisms for nonthermally producing ultraheavy dark matter \cite{Azatov:2021ifm,Baldes:2022oev,Ai:2024ikj,Giudice:2024tcp} or the baryon asymmetry of the Universe~\cite{Katz:2016adq,Baldes:2021vyz,Azatov:2021irb,Chun:2023ezg,Cataldi:2024pgt,Cataldi:2025nac} in various BSM setups.

Given such unique opportunities, it is worthwhile to ask whether such phenomena are possible with phase transitions associated with the only scalar field we have discovered so far, the Standard Model (SM) Higgs field. While the electroweak (EW) phase transition is a crossover transition in the SM, it can be made first-order in the presence of new physics, as known from various incarnations of electroweak baryogenesis \cite{Grojean:2004xa,Delaunay:2007wb,Espinosa:2007qk,Barger:2007im,Espinosa:2011ax,Kurup:2017dzf,Beniwal:2017eik,Ellis:2018mja,Bruggisser:2018mus,Ellis:2019oqb,Kozaczuk:2019pet} (see.\,e.g.\,\cite{Barrow:2022gsu} for a recent review). However, such modifications generally require thermal corrections generated by the presence of a thermal bath, and interactions between the bath and the expanding Higgs bubbles produce a frictional force that limits the bubble walls to non-relativistic speeds (which is, indeed, a crucial ingredient for electroweak baryogenesis).  

However, there exist other variations (beyond such electroweak-baryogenesis-type FOPTs) that can produce UR Higgs bubbles. One possibility is to look for BSM extensions that can realize a ``supercooled" EW FOPT that can occur at very low temperature. Another direction is to consider transitions associated with a SM Higgs minimum other than the EW vacuum. According to current measurements of SM parameters, the SM Higgs potential becomes unstable at high scales beyond $\sim 10^{11}$ GeV, and another Higgs minimum appears at large field values, carrying interesting cosmological implications \cite{Ellis:2009tp,Espinosa:2007qp,Elias-Miro:2011sqh,Kobakhidze:2013tn,Hook:2014uia,Kobakhidze:2014xda,Shkerin:2015exa,Kearney:2015vba,Espinosa:2015qea,East:2016anr,Grobov:2016llk,Kohri:2016qqv,Markkanen:2018pdo,DeLuca:2022cus,Espinosa:2017sgp,Espinosa:2018euj,Espinosa:2018eve,Hook:2019zxa,Shakya:2023zvs}. In particular, it was recently pointed out \cite{Shakya:2025mdh} that the most natural initial configuration of the SM Higgs field at the beginning of our Universe involves a FOPT from this higher minimum towards the EW vacuum, where the Higgs bubbles reach UR speeds despite significant interactions with the surrounding plasma due to modified dynamics associated with the symmetry-restoring nature of the transition \cite{Long:2025qoh}. As we will discuss in this paper, the existence of this higher Higgs vacuum and/or the possibility of suppressing thermal friction effects in symmetry-restoring transitions can lead to several novel configurations that produce UR Higgs vacuum bubbles.

In such frameworks, the collisions of UR Higgs bubbles can act as high energy colliders capable of producing particles with masses and energies far higher than any temperature reached in our cosmic history. There is no shortage of well-motivated new physics that couples to the SM Higgs field at high scales; $\textit{any}$ such state -- heavy superpartners, Kaluza-Klein excitations, new states associated with Grand Unified Theories (GUTs) -- can be produced by these cosmic Higgs colliders. Such configurations also provide unique opportunities to address fundamental problems in cosmology. In this paper, we will consider two such applications.

 The first is the nonthermal production of ultraheavy Higgs portal dark matter (DM) \cite{Arcadi:2019lka}. An attractive feature of such DM candidates is that they necessarily have sizable interactions with the SM, making them accessible to a multitude of experimental searches. As another application, we will consider the production of heavy Majorana right-handed neutrinos (RHNs), which are one of the most well-motivated BSM particles expected to couple to the Higgs in the most straightforward realization of neutrino masses, the type-I seesaw mechanism. The decay of such RHNs provides an elegant way to produce the observed baryon asymmetry of the Universe via leptogenesis \cite{Fukugita:1986hr}.

Section \ref{sec:scenarios} discusses the conditions for producing UR bubbles, and presents several specific scenarios that achieve such configurations with the SM Higgs field. Section \ref{sec:pheno} discusses the production of high mass/energy particles from such Higgs bubble collisions, and presents two cosmological applications: production of ultraheavy Higgs-portal dark matter and the corresponding direct and indirect detection signals (Sec.\,\ref{subsec:DM}), and production of heavy RHNs leading to leptogenesis (Sec.\,\ref{subsec:leptogenesis}). Some concluding thoughts and discussions are presented in Section \ref{sec:summary}. The main results of the paper are concisely summarized in Table \ref{table}.

\section{Scenarios with Relativistic Higgs Bubbles}
\label{sec:scenarios}

We are interested in a FOPT of the SM Higgs field from a metastable, false vacuum characterized by a vacuum expectation value (vev) $v_i$ to a stable vacuum
configuration with vev $v_f$. The nucleated bubbles of the new vacuum expand due to the latent vacuum energy released in the transition, which we parameterize as $\Delta V=c_v \Delta v^4$, where $\Delta v= v_i-v_f$. 

In the presence of a thermal plasma with temperature $T$, this outward vacuum pressure is countered at leading order by thermal pressure due to particles entering the bubbles, which evaluates to \cite{Bodeker:2009qy} $\mathcal{P}_{\text{LO}}\approx \frac{1}{24}\Delta m^2 \,T^2$ for a single relativistic particle species in the bath, with $\Delta m^2=m_{f}^2-m_{i}^2$, with $i$ and $f$ subscripts denoting the initial and final vacua. The next-to-leading order pressure contribution comes from transition radiation, the emission of vector bosons as particles in the thermal bath traverse the bubble walls. In the limit of $\gamma\gg1$, where $\gamma$ is the Lorentz boost factor of the bubble wall in the plasma frame, this is given by \cite{Bodeker:2017cim} (see also
\cite{Espinosa:2010hh,Dorsch:2018pat,Hoche:2020ysm,Gouttenoire:2021kjv,Ai:2024shx,Long:2024sqg,Azatov:2024auq,Ai:2025bjw}) $\mathcal{P}_{\text{NLO}}\sim g^2 \gamma |\Delta m_V| T^3$, where $g$ is the gauge coupling.  However, $\mathcal{P}_{\text{NLO}}$ can have a modified form, and can in particular be $\textit{negative}$, for intermediate $\gamma$ in symmetry-restoring transitions, where $v_f<v_i$ \cite{Long:2025qoh}.  In standard BSM models of electroweak FOPTs (employed for EW baryogenesis), a thermal SM plasma is present, and $v_i=0$ and $v_f\approx 100$ GeV. In such configurations, $\mathcal{P}_{\text{LO}}$ is positive and saturates $\Delta V$, so that the Higgs bubble walls only achieve nonrelativistic terminal velocities (with $\gamma\sim 1$). 

Based on the above discussion, there are essentially two avenues for achieving UR Higgs bubbles:

\textit{ Zero temperature FOPT:}  In the absence of a thermal plasma that interacts with the Higgs bubble walls, the aforementioned pressure contributions do not exist, and the released vacuum energy can drive the expanding Higgs bubbles to UR speeds. Note that this does not mean that the Universe must be vacuum dominated;  a radiation bath (e.g.\,corresponding to a dark sector) can exist as long as it does not interact appreciably with the Higgs bubbles. 

\textit{Symmetry-restoring FOPT:} If the transition involves $v_f<v_i$, then $\mathcal{P}_{\text{LO}}\sim \Delta m^2 \,T^2$ is negative (see discussions in \cite{Buen-Abad:2023hex,Azatov:2024auq,Long:2025qoh}) since the SM particle masses are proportional to the Higgs vev, resulting in $\Delta m^2<0$. For intermediate $\gamma$ values, $\mathcal{P}_{\text{NLO}}$ is also negative \cite{Long:2025qoh}, and nothing counteracts the accelerating bubble walls.  As $\gamma$ increases, the NLO pressure becomes positive and significant, eventually forcing a terminal velocity that can nevertheless be ultrarelativistic.  

For FOPTs that do not involve gauge bosons changing in mass, $\mathcal{P}_{\text{NLO}}=0$, and a third option exists: a thermal transition with $\Delta V >\mathcal{P}_{\text{LO}}$. This option is not viable for the SM Higgs, since the $W$ and $Z$ gauge boson masses are tied to the Higgs vev, producing NLO pressure that will eventually overcome vacuum pressure. 

We now discuss several specific FOPT configurations for the SM Higgs field that satisfy the above criteria, enabling UR Higgs bubbles. The parameters most relevant for the phenomenology of Higgs bubble collisions are (i) the strength of the phase transition $\alpha$, defined as the ratio of the vacuum energy density $\Delta V$ to the total energy density in the Universe, (ii) the inverse duration of the phase transition $\beta$, expressed as a dimensionless parameter $\beta/H$, where $H$ is the Hubble rate at the time of the transition, (iii) the temperature $T_*$ of the thermal bath after the transition completes, and (iv) the Lorentz boost factor $\gamma$ of the bubble walls at the time of collision.

\subsection{Supercooled Electroweak Transition (scEW)}
\label{sec:cold}

For the SM Higgs to undergo an electroweak symmetry-breaking FOPT at essentially zero temperature (i.e.\,in the absence of a SM thermal bath), it must be coupled to another field that drives the FOPT. Such configurations can be realized in several models. For example, in composite Higgs models, a light dilaton can drive a first-order confinement phase transition of new strong dynamics; this confinement process produces a composite Higgs, simultaneously triggering an EW phase transition, which is therefore also first-order \cite{Konstandin:2011dr,Konstandin:2011ds,Servant:2014bla,vonHarling:2016vhf,Bruggisser:2018mrt,Bruggisser:2022rdm,Bruggisser:2022ofg,Bhusal:2025lvm}. Another possibility is a singlet-scalar extension of the SM, where a real, gauge-singlet scalar field that couples to the SM via a Higgs portal interaction drives this behavior \cite{Kurup:2017dzf} (see also \cite{Kobakhidze:2017mru}). Such transitions can be supercooled, i.e.\,become efficient only after a significant drop in temperature removes the barrier separating the two phases, and the FOPT proceeds at essentially zero temperature. Consequently, the Higgs bubbles encounter negligible thermal friction and reach UR speeds. 

The exact details of the triggered FOPT are model-dependent, but the parameters relevant for the phenomenology of the ensuing UR Higgs bubble collisions are largely independent of such details. For supercooled transitions, we have $\alpha\sim1$. Studies of FOPTs in composite Higgs models \cite{Bruggisser:2022rdm} suggest that $\beta/H\approx 100$ and $T_*\approx 100$ GeV (note that we must have $T_*< 130$ GeV, otherwise EW symmetry will be restored). Likewise, the size of a critical nucleated bubble (relevant for calculating $\gamma$ at collision) is expected to be $R_c\sim m_h\approx 100$ GeV \cite{Bhusal:2025lvm}.

\subsection{Thermal Symmetry-Restoring Transition (tSRT)}
\label{sec:wrongtunnel}

Based on current measurements of the SM parameters, the SM Higgs quartic coupling runs to negative values at large field values, giving rise to a deeper Higgs minimum. Assuming the potential is stabilized by new physics coupling to the Higgs at some ultraviolet scale $\Lambda_{UV}$,  the finite temperature Higgs potential can be approximated (ignoring the mass term, which is negligible at large field values) as
\be
V_h(h)\approx-\frac{0.04}{4\pi^2}\text{ln}\left(\!\frac{h^2}{h_{\text{max}}^2\sqrt{e}}\!\right)h^4+\frac{h^6}{\Lambda_{UV}^2}+0.06 \,h^2\, T^2\, e^{-h/(2\pi T)}
\label{eq:potential}
\ee
The first term approximates the running of the Higgs quartic coupling in the SM \cite{Espinosa:2015qea}, which causes the potential to reach a local maximum at $h_{\text{max}}\approx 4\times 10^{11}$ GeV. The second term is an ansatz to capture the effects of new physics coupling to the Higgs at scale $\Lambda_{UV}$, which creates a (zero-temperature) deeper minimum at $v_{UV}\approx 0.04\, \Lambda_{UV}$. The final term represents thermal corrections from a SM plasma with temperature $T$ \cite{Espinosa:2017sgp,Franciolini:2018ebs,Hook:2019zxa}. 

If the SM Higgs field sits at $v_{UV}$, as would be natural from the point of view of initial conditions, a large reheat temperature after inflation can drive the Higgs to the EW minimum. This transition is likely first-order \cite{Shakya:2025mdh}, since the barrier separating the two minima persists as the deeper vacuum gets lifted above the EW vacuum by the thermal effects but becomes vanishingly small as the temperature rises further, and tunneling inevitably becomes efficient. This symmetry-restoring transition was calculated to take place from $v_i\approx 0.033 \Lambda_{UV}$ to $v_f\approx 0.02 \Lambda_{UV}$, with $\alpha\!\lesssim\!10^{-3},~\beta/H\sim7000,~T_*\approx 0.009\, \Lambda_{UV}$, and $R_c\approx 2000/\Lambda_{UV}$ for reasonable choices of reheating parameters \cite{Shakya:2025mdh}. We will henceforth fix $\Lambda_{UV}=10^{13}$ GeV for this tSRT scenario.

Note that since $v_f\!<\!v_i$, the SM particles lose mass across the transition, hence $\mathcal{P}_{\text{LO}}\approx \frac{1}{24}\Delta m^2 \,T^2$ is negative and aids the bubbles in accelerating. The NLO pressure is also negative at intermediate $\gamma$, becoming positive and enforcing a terminal velocity only when $\gamma$ grows somewhat large \cite{Long:2025qoh}. For the SM Higgs case being discussed here, the terminal boost factor of the bubble walls is estimated to be reached at $\gamma\lesssim 100$. 

\subsection{Symmetry-Restoring EW Transition (SR-EW)}
\label{sec:srelectroweak}

Inspired by the previous subsection, one can wonder whether a symmetry-restoring transition that ``inverts" a first-order electroweak transition might be possible during reheating. Models that modify the SM Higgs potential with new physics in order to engineer a FOPT \cite{Grojean:2004xa,Delaunay:2007wb,Espinosa:2007qk,Barger:2007im,Espinosa:2011ax,Kurup:2017dzf,Beniwal:2017eik,Ellis:2018mja,Bruggisser:2018mus,Ellis:2019oqb,Kozaczuk:2019pet} do so by creating thermal effects that produce a barrier between the two vacua when the EW vacuum becomes energetically favorable, so that the Higgs tunnels from the origin to the EW minimum. When the Universe reheats after inflation ends, this process could in principle be made to run in reverse: when the EW minimum is lifted above the symmetric vacuum at the origin, if a thermal barrier persists, this reverse, EW symmetry-restoring transition from the EW vacuum to the origin can be first-order. This has not been explicitly demonstrated in any paper in the literature, but should at least be considered conceivable (and has been alluded to in \cite{Barni:2025ced}). Note that this is more challenging that the symmetry-restoring transition discussed in the previous subsection (from a deeper Higgs mimimum at large field values beyond the EW minimum), where the barrier between the two vacua exists due to the RGE running of the Higgs quartic coupling in the SM rather than due to thermal effects. 

While building an explicit working model to this effect is beyond the scope of this paper, one can make reasonable estimates for the parameters of this would-be FOPT. Since this is an EW symmetry-restoring transition, one must have $T_* > 130$ GeV. Since thermal effects are responsible for driving the transition, we expect $\Delta V \sim T_*^4$, hence $\alpha< 10^{-2}$. For FOPTs driven by thermal effects, one expects $\beta/H\gtrsim 10^{3}$. Finally, the thermal pressure is determined by SM particles losing their mass across the transition, hence must be similar to the symmetry-restoring case discussed in the previous subsection, giving a terminal Lorentz factor $\gamma\lesssim 100$.

\subsection{Symmetry-Restoring Vacuum Transition (vSRT)}
\label{sec:quantumtunnel}

As a final variation, consider the case where the high field value minimum of the Higgs (see tSRT subsection) has higher potential energy than the EW minimum. This can occur if the stabilizing correction from new physics at $\Lambda_{UV}$ comes into effect immediately after RGE effects cause the Higgs potential to turn over, but before the potential reaches negative values, i.e.\,for $\Lambda_{UV}\gtrsim h_{\text{max}}$ in Eq.\,\ref{eq:potential}.  This is, admittedly, a fine-tuned scenario, but remains a possibility\,\footnote{A metastable, positive energy higher Higgs vacuum has been studied in earlier papers \cite{Masina:2011un,Masina:2011aa,Masina:2012yd} as a possible setting for inflation.} as the form of the Higgs potential at large field values is unknown.  Since this is a qualitatively different configuration capable of producing UR Higgs bubbles, it is worth considering for the purposes of this paper. 

As the higher minimum sits at a higher potential, thermal corrections are not needed to trigger the FOPT from this vacuum towards the origin. If a radiation bath is completely absent and the Universe is vacuum energy dominated, additional mechanisms are needed to avoid the graceful exit problem encountered in old inflation (see discussions in \cite{Masina:2011un,Masina:2011aa,Masina:2012yd}). These complications can be avoided by assuming that a radiation bath exists with temperature $T_{\text{min}}>\Delta V_h^{1/4}$ (where $\Delta V_h \approx 4.5\times 10^{38}$ GeV is the height of the Higgs potential barrier\,\cite{Espinosa:2015qea}) but corresponds to a secluded sector that couples negligibly to the Higgs, so that there are no thermal corrections to the zero-temperature Higgs potential. 

We require that the Higgs field tunnel out of the higher minimum into the EW vacuum before the temperature drops below $T_{min}$ in order to avoid transitioning into a vacuum-dominated Universe. This is accomplished if the tunneling probability per Hubble volume per Hubble time is greater than $1$, i.e.\,\cite{Fairbairn:2019xog} 
\be
\Gamma \approx v_{UV}^4 \left(\frac{S_4}{2\pi}\right)^2 e^{-S_4} > H^4 \approx \frac{T_{min}^8}{M_P^4}\,,
\ee
where $S_4$ is the $O(4)$-symmetric bounce action, and $M_P$ is the Planck scale. 

Since the thin-wall approximation is not applicable to the SM Higgs potential, $S_4$ must be evaluated numerically; using the numerical package {\tt{FindBounce}} \cite{Guada:2020xnz,Guada:2018jek}, we find that a FOPT occurs as desired if $\Lambda_{UV}\lesssim 127\, h_\text{max}$. With this, the position of the minimum almost coincides with the position of the peak of the barrier, $v_{UV}\approx h_\text{max}$, hence the barrier becomes vanishingly small, resulting in efficient tunneling. Note that since the Higgs field in the metastable minimum is massive, $m_h^2\sim v_{UV}^2 \gg H^2$, quantum fluctuations during an earlier period of inflation that can knock the field into the EW minimum over this vanishingly small potential barrier are exponentially suppressed, enabling it to survive through inflation in this configuration.  

If $T_*\approx T_{min}$, we can have $\alpha\sim 1$; for a quantum transition (where the shape of the potential does not change appreciably as the temperature drops), one generally has $\beta/H\sim 10$ \cite{Fairbairn:2019xog}; and the size of the critical bubble is expected to be $\mathcal{O}(10\,v_{UV}^{-1})$.

\section{Phenomenology}
\label{sec:pheno}

Having established in the previous section that there exist several early Universe configurations that can produce UR Higgs bubbles from a FOPT associated with the SM Higgs, we now explore various phenomenological opportunities that this can give rise to. 

\subsection{Particle Production from Bubble Collisions}

When thermal friction is negligible (as in the scEW and vSRT cases above), the boost factor $\gamma$ of the bubble grows linearly with its size, $\gamma\approx R/R_c$. Since $R\approx \beta^{-1}$, the typical bubble size when the bubbles collide, is of the order of the inverse Hubble scale, this gain can be sizable. The bubble wall thickness $\l_w\sim R_c$ gets Lorentz boosted to $R_c/\gamma$ at collision; this represents the (inverse) energy per unit area of the bubble wall at collision, and sets the maximum energy scale accessible to these bubble collisions, which can be estimated as \cite{Giudice:2024tcp}
\be
E_\text{max}\sim 2\gamma/R_c\approx \frac{0.03}{\beta/H}\frac{1}{T_*^2 R_c^2} M_P\,.
\ee
A rigorous calculation of particle production from such collisions makes use of the effective action formalism \cite{Watkins:1991zt,Konstandin:2011ds,Falkowski:2012fb,Shakya:2023kjf,Mansour:2023fwj,Giudice:2024tcp}, and amounts to calculating the decays of the distribution of off-shell excitations of the FOPT field, which has a UV cutoff $\sim E_\text{max}$. 
 
 On the other hand, for symmetry-restoring phase transitions in the presence of a thermal plasma (tSRT and SR-EW above), the Higgs bubble walls are expected to reach a terminal velocity with $\gamma\lesssim 100$, and the bubble collisions have a (relatively) lower energy reach $E_\text{max}\sim 200/R_c$. 

We will now discuss two cosmological applications of this enhanced energy reach of UR Higgs bubble collisions. 

\subsection{Higgs Portal Dark Matter}
\label{subsec:DM}

We now consider the production of scalar Higgs portal dark matter (see e.g.\,\cite{Arcadi:2019lka} for a review), a well-motivated dark matter setup that makes use of the only renormalizable coupling allowed between the SM and a stable DM candidate. The setup consists of a scalar  DM candidate $\chi_s$  with mass $m_{\chi_s}$ that couples to the Higgs via the Higgs portal coupling $\frac{1}{4}\lambda_s H^\dagger H \chi_s^2$. Note that if $m_{\chi_s}^2\!\gg m_h^2$, this coupling can produce radiative contributions that can lift the Higgs mass to the DM scale, requiring significant fine-tuning to keep the Higgs light; we will ignore this well-known hierarchy problem. Similarly, it can also produce corrections to the Higgs potential, but this is only expected to become appreciable at scales comparable to the DM mass scale. 

When UR Higgs bubbles collide, the off-shell Higgs excitations $h^*$ can decay into DM via the decay processes $h^*\to\chi_s^2,~ h\chi_s^2$ enabled by the portal coupling above. A comprehensive study of DM production at various stages of a generic FOPT has been performed in \cite{Giudice:2024tcp}. Using the results here, the DM relic abundance from Higgs bubble collisions can be written as
\bea
\Omega_\chi h^2&\approx& 0.1\,\frac{\beta/H}{10}\left(\frac{\alpha}{(1+\alpha)100\, c_V}\right)^{1/4}\frac{\lambda_s^2\,m_{\chi_s}\,v_h}{(24~\mathrm{TeV})^2}\nonumber\\
&&\times \,\left[\frac{v_h^2}{m_{\chi_s}^2}+\frac{1}{16\pi^2}\,\ln\left(\frac{E_\text{max}}{(2m_{\chi_s}+m_h)}\right)\right]\,.
\label{eq:scalarabundance}
\eea

Since we have estimates for all the parameters in the four FOPT scenarios discussed in the previous section (see Table \ref{table} for a convenient overview), we can invert the above relation to obtain the value of $\lambda_s$ that produces the observed DM relic density for a given DM mass. 

\begin{figure}[t]
\begin{center}
 \includegraphics[width=.99\linewidth]{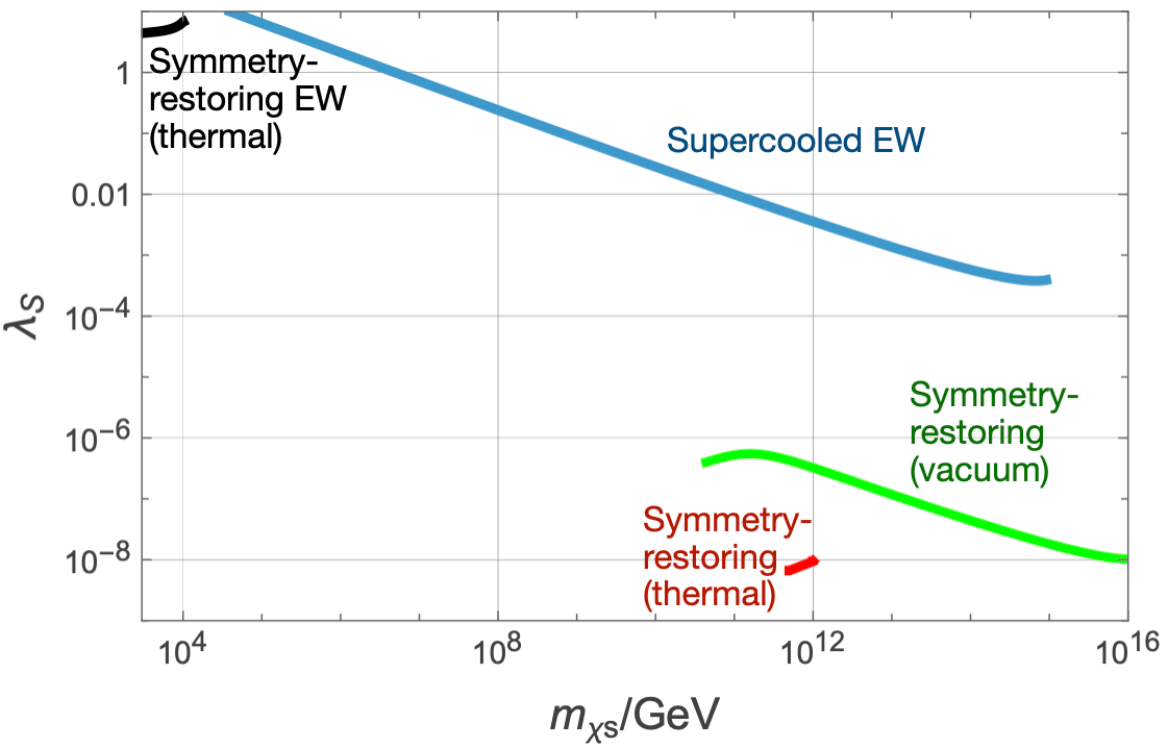}
  \end{center}\vspace{-4mm}
\caption{\textbf{Dark Matter Parameter Space:} Parameter space for scalar Higgs portal dark matter produced via ultrarelativistic Higgs bubble collisions in the four scenarios discussed in Section \ref{sec:scenarios} (see also Table \ref{table}).}
\label{fig:param}
\end{figure}

We show the dark matter mass range for which Higgs bubble collisions can produce the observed relic density and the associated values of the portal coupling for the four scenarios discussed in the previous section in Figure \,\ref{fig:param} (see also Table \ref{table}). As expected, the transitions that take place in the absence of a friction-inducing SM plasma (scEW and vSRT) feature runaway bubbles, enabling the production of DM with masses many orders of magnitude above the transition scale, up to $\mathcal{O}(10^{15})$ GeV, unleashing the full potential of a cosmic collider. 
In contrast, the two thermally induced symmetry-restoring scenarios (SR-EW and tSRT) where the bubbles reach a terminal velocity with $\gamma\approx 100$ have limited reach, only producing DM with masses one to two orders of magnitude above the transition scale. For these latter (thermal) cases, Higgs particles in the bath can also up-scatter into DM particles when they cross into the bubbles \cite{Azatov:2020ufh,Azatov:2021ifm,Azatov:2022tii,Baldes:2022oev}, providing another production mechanism for DM; this channel is either inactive or subdominant for the cases plotted in the figure. 

It is also worth noting that the EW-scale transitions (scEW and SR-EW) feature $\mathcal{O}(1)$ couplings, while the high scale transitions (tSRT and vSRT) feature significantly smaller couplings $\lambda_S\!<\!\mathcal{O}(10^{-6})$. This is because the high scale transitions occur at higher temperatures, when the Hubble radius is significantly smaller: this results in more bubble collisions per unit volume, resulting in greater DM production, therefore requiring smaller couplings. For the EW-scale transitions, DM can thermalize after production and undergo subsequent freezeout; to avoid this complication, we limit our study to $m_{\chi_S}\!>\!3$ TeV for these cases, so that $m_{\chi_S}/T\!>\!30$ and DM remains out of equilibrium. We also ignore the regime $\lambda_S\!>\!10$, where the theory is clearly not perturbative. The SR-EW case does not yield any suitable DM candidates for $\beta/H<2\times 10^4$; the shown curve (black) corresponds to the choice $\beta/H=10^5$.   For the high scale transitions with smaller couplings, freeze-in production from the thermal plasma can dominate if the DM mass is sufficiently close to the temperature of the bath \cite{McDonald:2001vt,Hall:2009bx,Elahi:2014fsa,Frangipane:2021rtf}; in the plot, we terminate these curves (to the left) where this occurs. 

\begin{figure}[t]
\begin{center}
 \includegraphics[width=.99\linewidth]{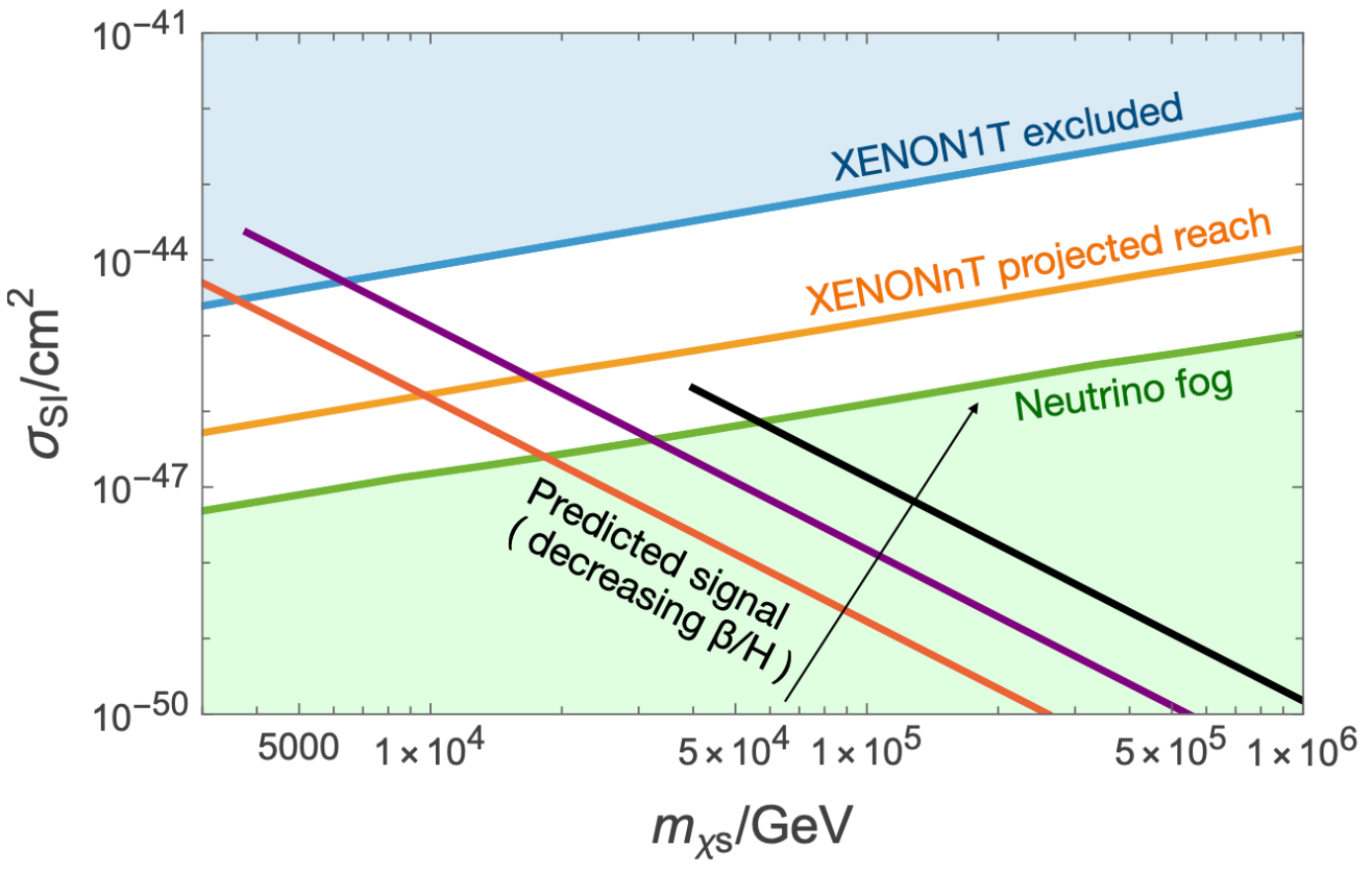}
  \end{center}\vspace{-4mm}
\caption{\textbf{Direct Detection:} Predicted spin-independent direct detection cross section for dark matter produced from the scEW transition for (top to bottom) $\beta/H=100,1000,10000$, truncated to the left when $\lambda_S\!>\!10$, with current exclusion limits from XENON1T and the projected reach of XENONnT (20t$\times$y)  (from \cite{Aprile:2024fkg}).}
\label{fig:dd}
\end{figure}

For the EW-scale transitions, the $\mathcal{O}(1)$ DM-Higgs portal couplings imply sizable indirect and direct detection signals, leading to excellent observational prospects. The spin-independent scattering cross-section with nuclei in Higgs portal models can be written as\,\cite{Arcadi:2019lka} 
\be
\sigma_{SI}\approx\frac{0.1 \lambda_s^2}{16\pi}\frac{m_\text{nuc}^4}{m_h^4}\frac{1}{m_{\chi_s}^2}\,,
\ee
where $m_\text{nuc}$ is the nucleon mass. Figure \ref{fig:dd} shows the predicted spin-independent direct detection cross section for DM produced from the scEW transition for various choices of the $\beta/H$ parameter. Also shown are current exclusion limits from XENON1T and the projected reach of XENONnT  (from \cite{Aprile:2024fkg}), as well as the neutrino floor/fog. As we can see, current and upcoming direct detection experiments can probe DM produced from Higgs bubble collisions in the scEW scenario up to $m_{DM}\approx\mathcal{O}(10)$ TeV.

\begin{figure}[t]
\begin{center}
 \includegraphics[width=.99\linewidth]{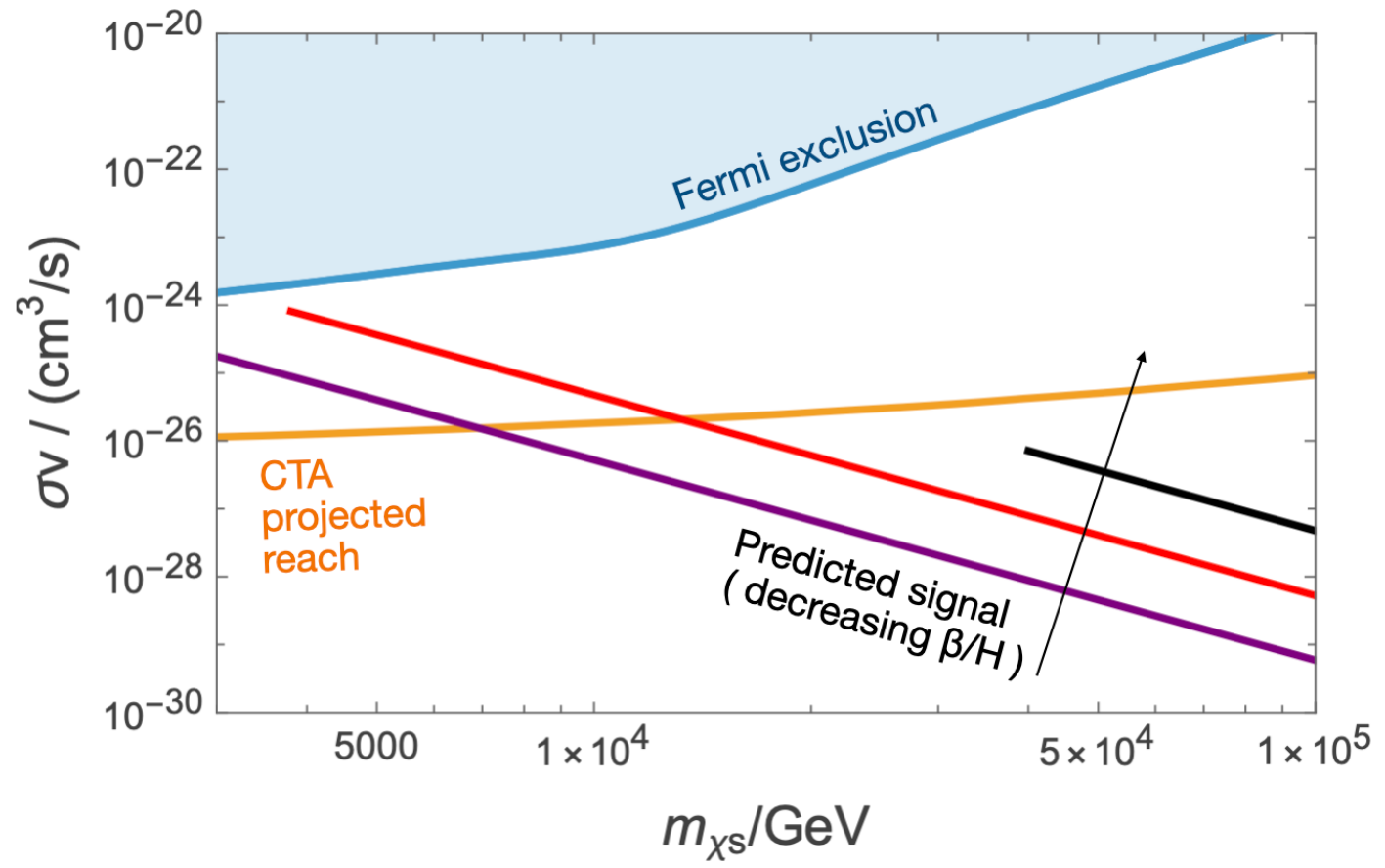}
  \end{center}\vspace{-4mm}
\caption{\textbf{Indirect Detection:} Predicted annihilation cross-section for dark matter produced from the scEW transition for $\beta/H=10,100,1000,10000$ (upper to lower), with current exclusion limits from Fermi (\cite{Elor:2015bho}) and the projected reach of the Cherenkov Telescope Array (CTA)  (\cite{CTA:2020qlo}).}
\label{fig:indd}
\end{figure}

For the same parameters, the corresponding indirect detection cross-section for DM annihilation into Higgs bosons is plotted in Figure \ref{fig:indd}. This s-wave cross section can be written  (for $m_{\chi_S}\gg m_h$) as\,\cite{Arcadi:2019lka} 
\be
\langle\sigma v\rangle_{\chi\chi\to hh}\approx\frac{\lambda_S^2}{32\pi m_{\chi_S}^2}\,.
\ee
The plot also shows exclusion limits from Fermi data \cite{Fermi-LAT:2015att} (as calculated for DM annihilation to Higgs bosons in \cite{Elor:2015bho}), and the projected reach of the Cherenkov Telescope Array (CTA)  \cite{CTA:2020qlo} (we use the reported limits for DM annihilation into $W$ bosons, which produces spectra similar to annihilation to Higgs bosons). It is worth noting that this present day annihilation cross-section can be larger than the canonical thermal target $\langle \sigma v\rangle\approx 2\times10^{-26} cm^3/s$, as DM was produced non-thermally from bubble collisions below its freezeout temperature\,\footnote{For a discussion of thermal histories that allow for  DM annihilation cross-sections larger than the canonical thermal target, see \cite{Puetter:2022ucx}.}. As with direct detection, we see that indirect detection offers observational prospects up to $m_{DM}\approx\mathcal{O}(10)$ TeV. 

Such Higgs-portal DM candidates with TeV scale masses and $\mathcal{O}(1)$ couplings are also expected to be produced copiously at high energy colliders (see discussions in \cite{Arcadi:2019lka}), and could produce observable signals at current and future colliders such as the high-luminosity Large Hadron Collider (LHC) and a Future Circular Collider (FCC), offering another complementary detection avenue. For ultraheavy DM (as produced in the scEW, tSRT, and vSRT frameworks), the traditional searches above have no scope, but other novel detection mechanisms might be relevant \cite{Carney:2022gse}.  
In all cases, stochastic gravitational waves produced from the FOPTs would serve as a complementary observable of the framework; the EW-scale transitions, in particular, are of interest for the upcoming LISA experiment  \cite{LISA:2017pwj,Caprini:2015zlo,Caprini:2019egz}.

\subsection{Leptogenesis}
\label{subsec:leptogenesis}

\begin{table*}[t]
    \centering
    \begin{tabular}{|c|c|c|c|c|c|c|c|c|} % Example with more columns
        \hline
        ~Scenario~ & ~Transition~ & ~$T_*$(SM)~ & ~$\alpha$~ & ~$\beta/H$~ & $c_V$~& ~$E_\text{max}$~ & ~Dark matter $(m_\chi)$~ & ~Leptogenesis $(M_N)$~ \\
        \hline
        ~scEW & $0\to v_{EW}$ & 100 GeV & $\sim 1$ & 100 & 0.1 & ~$4\times10^{15}$ GeV~ & ~$3$ TeV$-10^{15}$ GeV~ & no\\
       ~tSRT & ~$3\times 10^{11}$ GeV $\to 0$~ & $10^{11}$ GeV & ~$10^{-3}$~ & 7000 & ~$10^{-3}$~ & $2\times10^{12}$ GeV & ~$(5 - 10)\times 10^{11}$ GeV~ & $\sim 10^{12}$ GeV\\
        ~vSRT & ~$4.4\times 10^{10}$ GeV $\to 0$~ & ~$5\times10^{9}$ GeV~~& $\sim 1$ & 10 & $10^{-4}$ & ~$3\times10^{16}$ GeV~ & ~$\sim 10^{11}- 10^{16}$ GeV~& ~$4\times 10^{13}- 10^{16}$ GeV~\\
                ~SR-EW & $v_{EW}\to 0$ & 130 GeV & $10^{-2}$ & $10^5$ & $0.01$ & 25 TeV & $3-10$ TeV & no\\
        \hline
    \end{tabular}
      \caption{\textbf{Overview:} Scenarios (Section \ref{sec:scenarios}), relevant parameters, and phenomenology: dark matter (Section \ref{subsec:DM}) and leptogenesis (Section \ref{subsec:leptogenesis}).}
    \label{table}
\end{table*}

As another application, we briefly discuss the prospects of leptogenesis through the production of heavy right-handed neutrinos (RHNs) from Higgs bubble collisions. RHNs are one of the most well-motivated BSM extensions connected to the Higgs at high scales, providing the most straightforward explanation for the observed tiny neutrino masses through the type-I seesaw mechanism \cite{Minkowski:1977sc, PhysRevLett.44.912, Schechter:1980gr, Schechter:1981cv}, which consists of augmenting the SM Lagrangian with the terms
\be
\mathcal{L}\supset y_\nu \overline{L} H N +M_N \overline{N}^c N\,.
\label{eq:lagrangian}
\ee
The first term is a Dirac mass term between the SM lepton doublet $L$, the SM Higgs doublet $H$, and the SM gauge singlet RHN $N$, while the second term is a Majorana mass for the RHNs\,\footnote{The terms in Eq.\,\ref{eq:lagrangian} remain compatible with RHNs charged under BSM symmetries, see e.g.\,\cite{Roland:2014vba,Shakya:2018qzg,Roland:2016gli,Shakya:2015xnx}.}. In the presence of the above terms, electroweak symmetry breaking generates tiny masses for the SM neutrinos, $m_\nu\approx y_\nu^2 v_h^2/M_N$.  The natural RHN mass scale for obtaining $m_\nu\approx 0.05$ eV with $\mathcal{O}(1)$ Yukawa couplings $y_\nu$ is $M_N\approx 10^{14}$ GeV, close to the GUT scale; smaller RHN mass scales require correspondingly smaller couplings.  It is well-known that the existence of heavy RHNs facilitates leptogenesis \cite{Fukugita:1986hr}: RHN decays  produce a lepton asymmetry, which can get converted to a baryon asymmetry by sphaleron processes, producing the matter-antimatter asymmetry observed in the Universe. Successfully realizing this requires a high reheat temperature $T_R>M_N$ so that the RHNs are part of the thermal bath in the early Universe.

UR Higgs bubble collisions provide a different means of producing such GUT-scale RHNs: the off-shell Higgs excitations $h^*$ can decay as $h^*\to \nu+N$ even when $m_N\gg T_*$. RHN production from bubble collisions in generic FOPTs has been studied in \cite{Cataldi:2024pgt}, which provides an estimate for the RHN yield in our case
\be
\label{eq:neu_Y_neu}
Y_N=\frac{n_N}{s}\approx y_\nu^2\, \frac{\beta}{H}\left(\frac{\pi^2\alpha}{(1+\alpha) c_V}\right)^{\frac{1}{4}} \frac{v_h}{M_{P}}\,\text{ln}\left(\frac{E_{\text{max}}}{M_N}\right)\,.
\ee

The lepton (and subsequent baryon) asymmetry from the decay of such RHNs can then be calculated using the standard formalism \cite{Cataldi:2024pgt}. It should be noted that, in contrast to the canonical thermal leptogenesis scenarios, washout effects from the thermal plasma are completely negligible here since $m_N\!\gg\!T_*$. 

From these calculations, one finds that the EW-scale Higgs FOPTs (scEW and SR-EW frameworks) cannot produce enough RHNs to explain the observed baryon asymmetry, as the number density of bubbles is far too low at such low temperatures\,\footnote{Such EW-scale UR Higgs bubble collisions could generate the baryon asymmetry in other ways \cite{Konstandin:2011ds,Servant:2014bla,Bruggisser:2018mrt,Bruggisser:2022rdm,Bhusal:2025lvm}.}. The general study in \cite{Cataldi:2024pgt} found that the FOPT scale needs to be at least $10^{9}$ GeV for this mechanism to be viable. This is indeed the case for FOPTs associated with the higher field value Higgs vacuum (tSRT and vSRT frameworks), and these do contain viable parameter space for leptogenesis through RHNs produced from Higgs bubble collisions (see Table \ref{table}). 
\begin{itemize}
\item tSRT: Since the bubbles reach a terminal velocity with $\gamma\sim 100$, this scenario has very limited reach, and only a very narrow window around $M_N\approx 10^{12}$ GeV exists.
\item vSRT: This framework admits a significantly greater range spanning several orders of magnitude in RHN mass, from $4\times 10^{13}$ GeV (below this, the RHN yield becomes too low) to $10^{16}$ GeV.  The natural type-I seesaw scale with GUT-scale RHNs can therefore easily be incorporated in this setup. 
\end{itemize}

It should be mentioned that the RHNs produce corrections that modify the Higgs potential (see e.g.\,\cite{EliasMiro:2011aa}); however, such corrections only alter the potential above the RHN mass scale, hence are negligible in the above scenarios, where the RHN mass scale lies above the scale at which the FOPT occurs. 

If is also worth noting that the efficacy of RHN production from Higgs bubble collisions is hindered by the seesaw relation, which forces $y_\nu\!\ll\!1$ when $M_N\!\ll\! 10^{14}$ GeV. Other well-motivated particles that are free from such constraints could yield larger windows of viability; e.g.\,in supersymmetric theories, Higgs bubble collisions could produce heavy superpartners, whose decays can produce the baryon asymmetry \cite{Dimopoulos:1987rk,Claudson:1983js,Sorbello:2013xwa,Cui:2012jh,Cui:2013bta,Arcadi:2015ffa,Barbier:2004ez,Pierce:2019ozl,Grojean:2018fus}.

\section{Conclusions}
\label{sec:summary}

In this paper, we have discussed the production and applications of ultrarelativistic (UR) Higgs bubbles in a first-order phase transition (FOPT) associated with the Standard Model Higgs field. UR bubbles are not encountered in standard BSM-modified electroweak FOPTs (aimed at electroweak baryogenesis) due to significant interactions between Higgs bubbles and the surrounding SM plasma. However, there are two avenues that enable Higgs bubbles to reach UR speeds:  engineering a phase transition without a SM plasma, so that this thermal pressure is absent, or having a symmetry-restoring transition, where the plasma interaction with the bubbles gets modified. We discussed the realization of either case for transitions associated with the electroweak vacuum as well as a new Higgs vacuum at higher field values, whose existence is favored by current measurements of SM parameters, presenting four qualitatively different frameworks for producing UR Higgs bubbles (Section \ref{sec:scenarios}). In this paper, we have only provided a general discussion of these cases, and it will be worthwhile to explore, in particular, how the supercooled or symmetry-restoring electroweak transitions can be realized in specific BSM models. 

The existence of such configurations carry tremendous implications, as the collisions of such bubbles act as cosmic scale Higgs colliders capable of reaching energies far beyond the temperature of the plasma, close to the Planck scale, and can therefore probe new physics coupled to the Higgs at very high energies that would otherwise remain inaccessible to the early Universe as well as any of our experimental probes.

 In Section \ref{subsec:DM}, we discussed the nonthermal production of ultraheavy Higgs portal dark matter from the collisions of such UR Higgs bubbles, which can be accomplished to different extents in all four frameworks. We found that DM with mass up to $10^{16}$ GeV can be produced with the correct relic density, even when the temperature of the Universe never exceeds the electroweak scale, showcasing the high energy collider nature and efficiency of such bubble collisions. The Higgs portal interactions give rise to observable signals of such DM candidates in direct detection, indirect detection, as well as collider experiments, with signals larger than those associated with canonical thermal DM targets, expanding the parameter space of Higgs portal DM models. While we only discussed scalar DM, similar results should also hold for fermion or vector DM. 

We also discussed the possibility of leptogenesis through the production of GUT-scale right-handed neutrinos from UR Higgs bubble collisions (Section \ref{subsec:leptogenesis}). While this mechanism is not viable for EW scale transitions, we found that transitions associated with the high scale Higgs vacuum can successfully incorporate this mechanism as a viable explanation for the baryon asymmetry of the Universe.

These are simply indicative of the vast phenomenological opportunities opened by such cosmic Higgs colliders. Any high scale new physics that couples to the Higgs -- heavy superpartners, Kaluza-Klein excitations, additional states associated with Grand Unified Theories, extended Higgs or dark sectors -- can be produced from such Higgs bubble collisions in the early Universe, even if these states are significantly heavier than any temperature reached in our cosmic history. Such directions are worthy of further study, and will be pursued in future work.
It is worth noting that in all such cases, the FOPTs are also accompanied by the production of gravitational waves,  providing a unique complementary observable aspect of these phenomena.

%%%%%%%%%%%%%%%%%%%%%%%%%%%%%%%%%%%%%%%%%%%%%%%%%%
\section*{Acknowledgments} 
%%%%%%%%%%%%%%%%%%%%%%%%%%%%%%%%%%%%%%%%%%%%%%%%%%

The author is grateful to Giulio Barni and Geraldine Servant for helpful exchanges. 

\bibliography{higgsuniverse}{}

\end{document}